\documentclass[superscriptaddress,prl,twocolumn,amsmath,amssymb,floatfix,showpacs]{revtex4-1}

\usepackage{graphicx}
\usepackage{bm}
\usepackage{amssymb}
\usepackage{color}
\usepackage{epsfig}
\usepackage{subfigure}
\usepackage{amsmath}
\usepackage{siunitx}
\usepackage{hyperref}
\usepackage{bbold}
\hypersetup{
    pdfstartview={FitH},
   pdfpagemode={None}
   }
\newcommand{\be}{\begin{equation}}
\newcommand{\ee}{\end{equation}}
\begin{document}


\title { Universal frequency-dependent conduction of electron glasses}

\author{Ariel Amir}
\affiliation {Department of Physics, Harvard University, Cambridge, MA 02138, USA}

\begin {abstract}
Characterizing the frequency-dependent response of amorphous systems and glasses can provide important insights into their physics. Here, we study the response of an electron glass, where Coulomb interactions are important and have previously been shown to significantly modify the conductance and lead to memory effects and aging. We propose a model which allows us to take the interactions into account in a self-consistent way, and explore the frequency-dependent conduction at all frequencies. At low frequencies conduction occurs on the percolation backbone, and the model captures the variable-range-hopping behavior. At high frequencies conduction is dominated by localized clusters. Despite the difference in physical mechanisms at low and high frequency, we are able to scale all numerical data onto a single curve, using two parameters: the DC conduction and dielectric constant. The behavior follows the universal scaling that is experimentally observed for a large class of amorphous solids.

\end {abstract}


\pacs {72.20.Ee, 72.80.Ng}
 \maketitle

 Amorphous systems, and in particular glasses, still present major theoretical and experimental challenges. One of the most fascinating and appealing aspects of these systems is the fact that despite the complex, system-specific microscopic details of each particular glass, they share much in common with regards to the measured response of various observable such as their low-temperature properties \cite{low_t, moshe, leggett} and their aging behavior \cite{amir_PNAS}. A useful tool to probe such systems is the measurement of the response of the system to an external oscillating potential, namely, the frequency-dependent conductance. Remarkably, also for this observable there are robust and generic features which span very different experimental systems, from plastics to doped semiconductors, as described in various reviews \cite{jonscher, review_universal_ac, zvyagin}.

 Here, we focus, theoretically, on electron glasses, which have been the focus of extensive experimental and theoretical research in the last decade \cite{amir_review, eglass_book}, and where Coulomb interactions were reported to have striking effects on the statics \cite{ES_gap, efros} and dynamics \cite{zvi:2}. In contrast to previous works, here we are able to study the effect of the long-ranged Coulomb interactions on the universal scaling of the frequency-dependent conduction, in a self-consistent way, using the local mean-field approximation which has previously been shown to capture the slow relaxations \cite{zvi3, grenet1,grenet2,amir_glass}, aging \cite{amir_aging, zvi1, zvi2, grenet_new, grenet_2009}, noise spectrum \cite{amir_noise}, temperature-dependent DC conduction \cite{amir_vrh} and memory effects \cite{vaknin, yasmine} in electron glasses. We shall show that the typical crossover frequency from DC to AC behavior is governed by the DC conductance and dielectric constant. For example, for a sample with resistance of $1 G \si{\ohm}$ and linear dimension $1 \si{cm}$, the frequency would be of the order of $\omega_c \sim 1 \si{kHz}$, and inversely proportional to the dielectric constant.

\emph{Model definition.} We consider a system where conductance is carried by hopping between localized sites, with unscreened Coulomb interactions between the sites.  Within the local mean-field approach, interactions are taken on a mean-field level, yet the occupations are different at every site, in contrast to other mean-field approaches \cite{ioffe}.  We note that while this approximation captures the physical phenomena well, as in most mean-field approaches it is not expected to give exact coefficients \cite{bergli_mean_field}. Furthermore, the local mean-field approximation might miss certain features of the frequency-dependent response, namely, the enhancement of frequency dependent response at very high frequencies due to the negative correlations between closely spaced sites \cite{pollak_correlations}, phononless hopping \cite{phononless}, and the possible effect of many-particle transitions on the DC conductance \cite{eglass_book}. The equations of motion for the occupation numbers $n_i$ are \cite{amir_glass, amir_review}: $\frac{d n_i}{dt}= \sum_{j\neq i}\gamma_{ji}-\gamma_{ij}$,  with the rates $\gamma$ given by Fermi's golden rule \cite {efros}: \be \gamma_{ij} \sim |M_q|^2 \nu n_i (1-n_j)
e^{-\frac{r_{ij}}{\xi}}[N(|\Delta E|)+\theta(\Delta E)] \label {rates},\ee where the $\theta$-function expresses the spontaneous emission term, $M_q$ is the electron-phonon coupling matrix element and $\xi$ the localization length. Throughout the paper we set the Boltzmann constant $k_B \equiv 1$.

At equilibrium (or, in fact, any configuration stable with respect to single-particle hops), the occupations numbers follow Fermi-Dirac statistics \cite{amir_glass}.  While universality in the frequency response of systems with Fermi statistics has been shown for non-interacting models \cite{pasveer}, and the effect of interactions has been studied before using the pair approximation \cite{efros_ac, efros}, our approach will allow us to study the effects of strong Coulomb interactions on the response, at all frequencies. The frequency-dependent voltage exerted on the system field will result in oscillations in the occupation numbers and energies around their equilibrium statistics, and will lead to a finite current, which at any finite frequency will have both an in-phase and out-of-phase component.

\emph{Mapping to an electric circuit.} We will assume that the voltage is small, such that we are in the linear response regime. At any instance in time, we can define the instantaneous energy of a site as:

\begin{equation} E_i \equiv \epsilon_i + \sum_{j \neq i} n_j e^2 /r_{ij} +V^{ext}_i \label{energies},  \end{equation}
where $V^{ext}_i$ is the external potential due to the charges on the leads. For example, if the sample is placed between two conducting parallel plates in three-dimensions, a case we shall present results for later on, we have $V^{ext}_i \propto x_i$ where $x$ is the coordinate perpendicular to the plates.

We define a \emph{local} chemical potential, $\mu_i$, $n_i = \frac{1}{1+e^{(E_i-\mu)/T}}$, with $n_i$, $E_i$ and $\mu_i$ time-dependent. Denoting $n_i(t)=n^0_i+\delta n$, $E_i(t)=E^0_i+\delta E $ and $\mu_i(t)=\mu^0_i+\delta \mu $, we find that:
\be \delta n =- \frac{\delta E  -\delta \mu }{T}n^0_i (1-n^0_i). \label{deltan}\ee

At zero field, detailed balance assures us that $\gamma_{ij}=\gamma_{ji}$. This is broken by the voltage, and we can expand $\gamma_{ij}=\gamma^0_{ij}+\delta \gamma_{ij}$, with:
$ \delta \gamma_{ij}= \gamma^0_{ij}\left[\frac{\delta n_i}{n^0_i}-\frac{\delta n_j}{1-n^0_j}-\frac{ N(N+1)\delta E}{T[N+\theta(\Delta E)]} \right],$ where $\Delta E$ is the energy difference at equilibrium, $N$ is a Bose-Einstein function of it, and $\delta E = \delta E_i- \delta E_j$.

Calculating the total current between two sites, $I_{ij} \equiv \gamma_{ji}-\gamma_{ij}$, combined with Eq. (\ref{deltan}), gives:

\begin{equation} \frac{d n_i}{dt} = \sum_{j \neq i} I_{ij} = \sum_{j \neq i}  [\delta \mu_j-\delta \mu_i]/R_{ij} ,\label{currents}\end{equation}
where we have defined the Miller-Abrahams resistors \cite{miller_abrahams} as $R_{ij} \equiv \frac{T}{\gamma_{ij}}$. Combining Eqs. (\ref{deltan}) and (\ref{energies}), we can relate ${\delta n}$ and ${\delta \mu}$:

\begin{equation} \delta \mu_i = T\frac{\delta n_i}{n^0_i(1-n^0_i)}+\delta E_i = \beta_{ik} \delta n_k+V^{ext}_i , \label{connection}\end{equation}
with the matrix $\beta$ defined as:

\begin{equation} \beta_{ij}=\frac{T\delta_{ij}}{n^0_i(1-n^0_i)}+(1-\delta_{ij})e^2/r_{ij}. \label{beta}\end{equation}
For an oscillating voltage, we have $V^{ext}_i(t)=\hat{V}^{ext}_i e^{\mathfrak{i} \omega t}$. We thus seek sinusoidal solutions of the form $\delta n (t) = \delta \hat{n}e^{\mathfrak{i} \omega t}$, $\delta E (t) = \delta \hat{E}e^{\mathfrak{i} \omega t}$, $\delta \mu (t) = \delta \hat{\mu}e^{\mathfrak{i} \omega t}$. Using Eqs. (\ref{currents}) and (\ref{connection}), we obtain:

\begin{equation} \mathfrak{i} \omega \hat{\delta n_i} =\mathfrak{i} \omega \beta_{ij}^{-1} (\hat{\delta \mu_j}-\hat{V}^{ext}_j)=  \sum_{j \neq i}[\delta \mu_j-\delta \mu_i]/R_{ij}  \label {circuit1}\end{equation}
We can define a conductance matrix $\sigma_{ij} \equiv 1/R_{ij}$, choosing its diagonal elements such that the sum of every column vanishes. Eq. (\ref{circuit1}) then takes the form:

\be  [\mathfrak{i} \omega \beta^{-1}-  \sigma  ]\vec {\delta \mu}= \mathfrak{i} \omega \beta^{-1}\vec{V}^{ext} \label {circuit},\ee

For non-interacting systems, previous works have generalized the Miller-Abrahams resistor network to linear response at a finite frequency, and found that in addition to the resistances the circuit contains self-capacitances \cite{AC_0, AC_1, AC_2}. Eq. (\ref{circuit}) also incorporates the \emph{Coulomb interactions} at finite frequency, and is the main tool which we will use to find the frequency-dependent conductance.
The boundary conditions are dictated by the leads; Using Fermi's Golden rule and the Fermi-statistics in the leads, one finds that the rates to the leads are given by \cite{amir_vrh}: $I_{lead} = -[n_i-n_{FD}(E_i-\mu)]/\tau$, where $n_{FD}$ is the Fermi-Dirac distribution. Expanding around the equilibrium gives: $I_{lead} =-[\delta n_i + n^0_i(1-n^0_i)(\delta E_i - \delta {\mu})/T  ]/\tau ,$ with $\delta_{\mu}$ the change in the chemical potential of the lead, which drives the currents.
Using Eq. (\ref{deltan}), we find that:

\begin{equation} I_{lead} =-[(\delta \mu_i- \delta {\mu})n^0_i (1-n^0_i)/T]/\tau  \equiv -[\delta \mu_i- \delta {\mu}]/R_{i,L}, \label {leads}\end{equation}
where $R_{i,L}=T/\gamma_{i,L}$, reciprocal of the equilibrium hopping currents $\gamma_{i,L}=n^0_i (1-n^0_i)/\tau$, in complete analogy to the Miller-Abrahams resistor network of Eq. (\ref{currents}). It is convenient to define a vector of conductances to the left and right leads, $({\vec{\sigma}}_{L (R)})_i \equiv 1/R_{i,L (R)}$. Then, using Eqs. (\ref{circuit}) and (\ref{leads}) we find that the problem reduces to the following linear system of equations:

 \be [\mathfrak{i} \omega \beta^{-1}-  \tilde{\sigma}  ]\vec {\delta \mu}=  \mathfrak{i} \omega \beta^{-1}\vec{V}^{ext}+{\vec{\sigma}}_{L} \delta \mu_L +{\vec{\sigma}}_{R} \delta \mu_R \label {circuit2},\ee
where for $i \neq j$ we define $\tilde{\sigma}_{ij} \equiv \sigma_{ij}$, but the diagonal elements of $\tilde{\sigma}_{ij}$ also include the terms arising from the currents to the leads, so that $\tilde{\sigma}_{ii} \equiv \sigma_{ii}-\sigma_{i,L}-\sigma_{i,R}$.

This linear equation can be solved for given chemical potentials of the leads, and describes the internal distribution of local chemical potentials. From this it is straightforward to find the currents in the system $I_{jk} = \sigma_{jk}[\delta \mu_j-\delta \mu_k]$, and extract the conductivity (see the Supplementary Information for details). Due to the exponential dependence on distance between sites, the elements of the resistance matrix will be broadly distributed (a property which allows for treatment of the DC conductance in terms of percolation theory \cite{ambegaokar,Shklovskii1972,pollak_percolation}). Similarly, the Fermi-dirac distribution of occupation numbers will lead to a broad distribution of capacitances.

\emph{Connection to the system relaxation spectrum.-} Eq. (\ref{circuit2}) can be given a simple physical interpretation. Let us define $C = \beta^{-1}$. Then, Eq. (\ref{circuit2}) describes a system of entities with self-capacitance $C_{ii}$, connected with mutual capacitances $C_{ij}$ and resistors $R_{ij}$ \emph{in parallel } to them. The mutual capacitances arise directly from the Coulomb interactions, since without them the inverse of $\beta$ is a diagonal matrix. Without interactions, the diagonal elements are $n^0_i(1-n^0_i)/T$, which indeed has the physical significance of a local compressibility of non-interacting fermions, clarifying their contribution to the system self-capacitances. Another interesting observation regards a connection between the matrices $R$ and $C$ and the relaxation of the system back to equilibrium. It can be shown that for small perturbations this relaxation is described by the equation $\frac{d\vec{n}}{dt}=A \vec{n}$, with \cite{amir_noise}:

\begin{equation} A=\gamma \beta = \sigma \cdot  C^{-1} , \end{equation}
where $\gamma$ is the symmetric matrix of equilibrium hopping rates. The RHS is a generalization of the concept of an $RC$ time, with $C$ and $R$ replaced by \emph{matrices}, thus generating a whole spectrum of relaxation times, corresponding to the eigenvalues of the matrix $A$ \cite{amir_expmat}. Notice that $C^{-1}$ is the inverse of the capacitance matrix, but $\sigma_{ij}\equiv 1/R_{ij}$. We also note that in spite of the fact that $\beta$ and $\gamma$ are symmetric matrices, $A$ is non-hermitian. The fact that $A \beta^{-1}$ is nonetheless symmetric is a manifestation of Onsager's theorem \cite{onsager,amir_noise}, stating that the product of the relaxation matrix and the equal-time correlation matrix must be a symmetric matrix. The relation we have found between the relaxation and the frequency-dependent conduction connects two important physical properties of the system.

 After a transient, which can be very long in certain systems \cite{pollak}, the system will settle in this steady-state oscillatory solution, where the total current through the system can be found by solving the linear problem of capacitors and resistors.

\emph{Numerical solution of the circuit impedance and its universal scaling.} The self-consistent set of equations for the site energies can be found iteratively \cite{amir_glass}, upon which the matrices $\sigma$ and $C$ can be evaluated. The impedance at any given frequency can then be found by solving Eq. (\ref{circuit2}).

\begin{figure}[t]
 \includegraphics[width=0.45\linewidth]{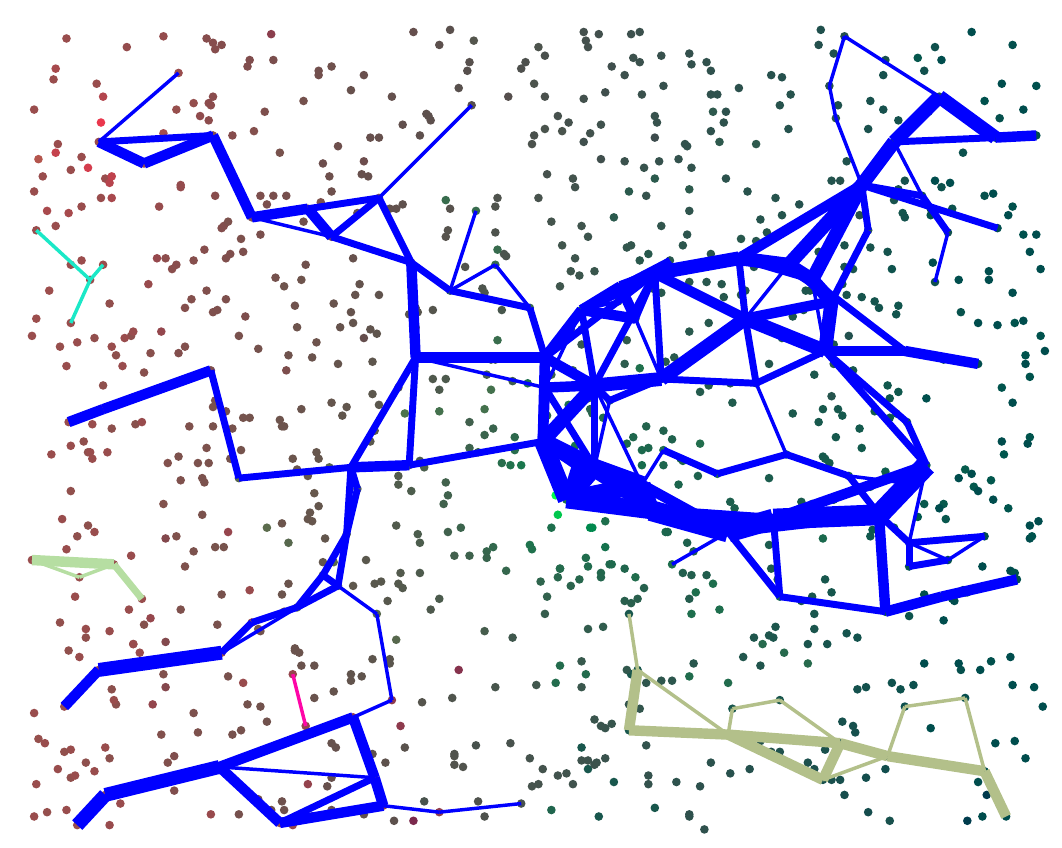} \includegraphics[width=0.45\linewidth]{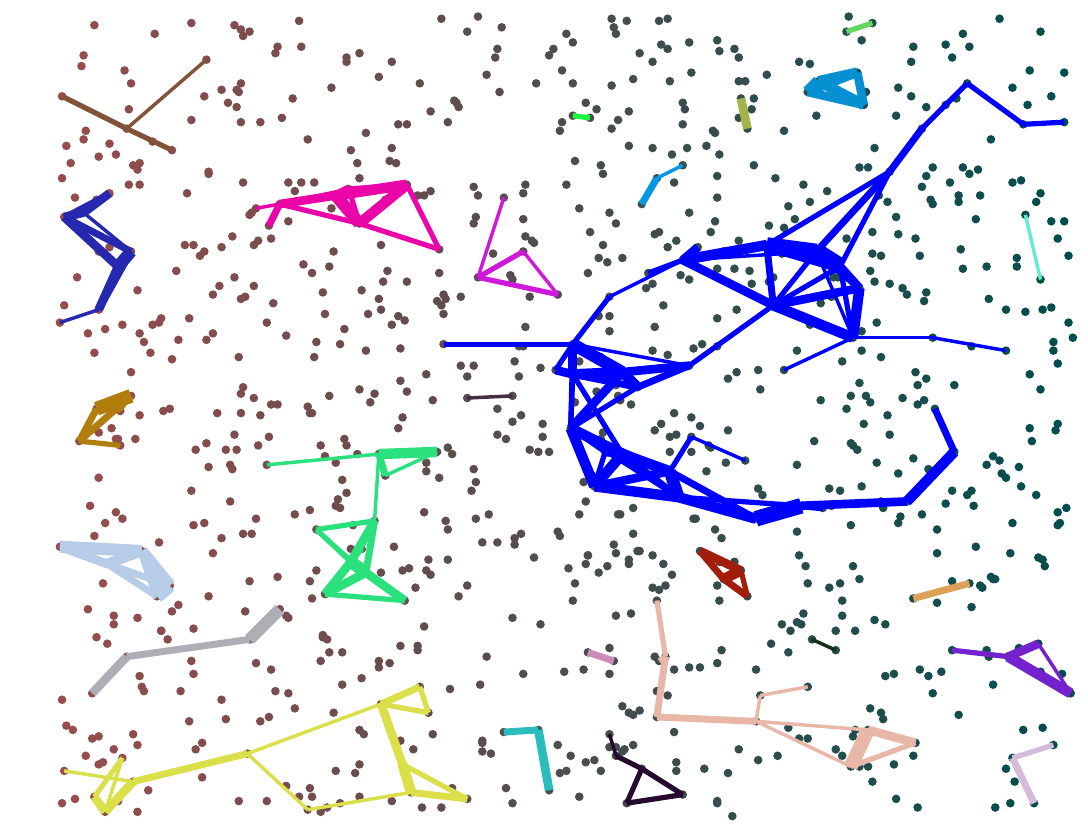} \\

\hspace {-3 cm} (a) \hspace {3.5 cm} (b)

\caption{The distribution of currents within a particular realization for $N=100$, $U \equiv \frac{e^2/{r_{nn}}}{W}=1$ and $T/W=0.05$ is shown, generated using the numerical procedure and model described in the main text for a 2d system, for two values of frequency ($\omega/\omega_c=10^{-10},10^{10}$ (one well below and the other well above the crossover frequency from the DC to the AC regime). Only currents larger than an arbitrarily chosen threshold value are shown, with the lines thickness indicating their value on a logarithmic scale. Each connectivity component is designated a separate color. (a) At low frequencies conduction occurs on the percolation backbone (b) Above $\omega_c$, localized clusters of well coupled electronic sites may contribute to the conduction. }
\label{pairs}
\end{figure}

While at low frequencies the conduction can be understood within the percolation picture \cite{ambegaokar,Shklovskii1972,pollak_percolation} (see Fig. \ref{pairs}a), at higher frequencies currents can occur in localized pairs or clusters, whose characteristic ``RC" time equals the reciprocal frequency of the AC driving, as shown in Fig. \ref{pairs}b. This is in line with previous work analyzing the contributions of localized pairs at high frequencies \cite{geballe,austin}. Note also that the localized clusters of Fig. \ref{pairs}b are correlated with the locations of the percolating backbone of Fig. \ref{pairs}a, suggesting that the conclusions of Ref. \cite{dyre3} can be extended also to strongly interacting systems. Summing the contributions of many such pairs at high frequencies explains the approximately linear dependence of conduction on frequency, until its saturation due to the finite size of the system, as explained in the Supplementary Information.

The transition from the localized pairs to the percolating network as one goes from high to low frequencies is also reminiscent of the structure of the eigenmodes of the conductance matrix, studied in Refs. \cite{amir_expmat,amir_prx} for the limit of the so-called $r$-hopping problem, where the temperature dependence of the matrix elements in omitted. There it is shown, using a combination of strong disorder renormalization group analysis and percolation theory, that the high frequency modes are due to localized pairs of points, while at lower frequencies these gradually hybridize leading to localized, finite clusters, which will suppress the frequency dependence \cite{pollak_clusters}. At a critical frequency a percolation transition takes place, and the eigenmodes become delocalized (two dimensions being a critical dimension for the transition). For the more general case where the matrix elements are temperature-dependent, a similar picture where pairs of site percolate at $\omega_c$ is discussed in Refs. \cite{hunt, hunt2}, where it is found that for the non-interacting case $\omega_c \sim \sigma_{DC} \propto  e^{-(T_0/T)^\alpha}$, with $\alpha=1/4$ the Mott variable-range-hopping exponent, for a non-interacting system.

 \begin{figure}[t!]
 \center{(a)}
\includegraphics[width=0.6\linewidth]{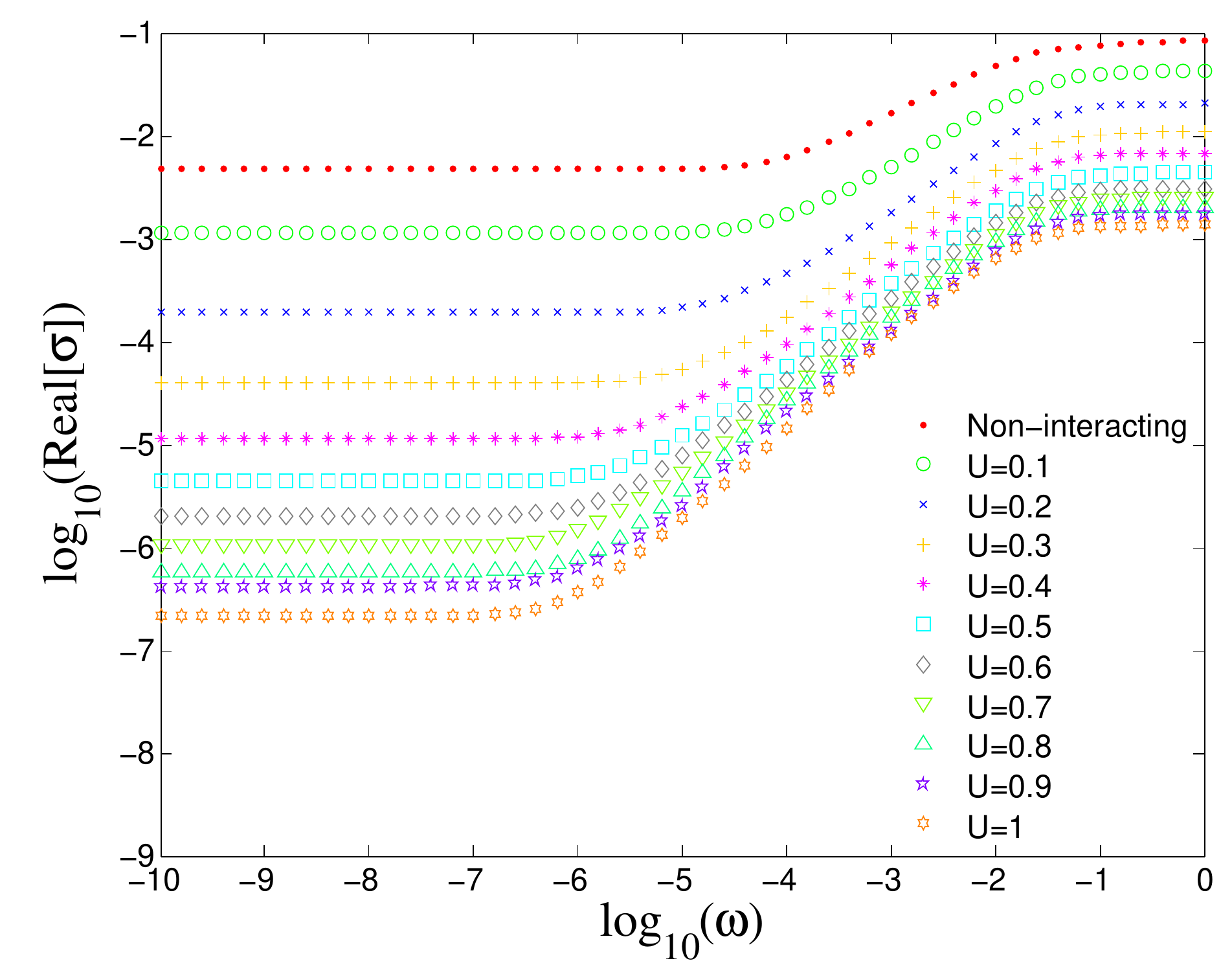}
\center{(b)}
\includegraphics[width=0.6\linewidth]{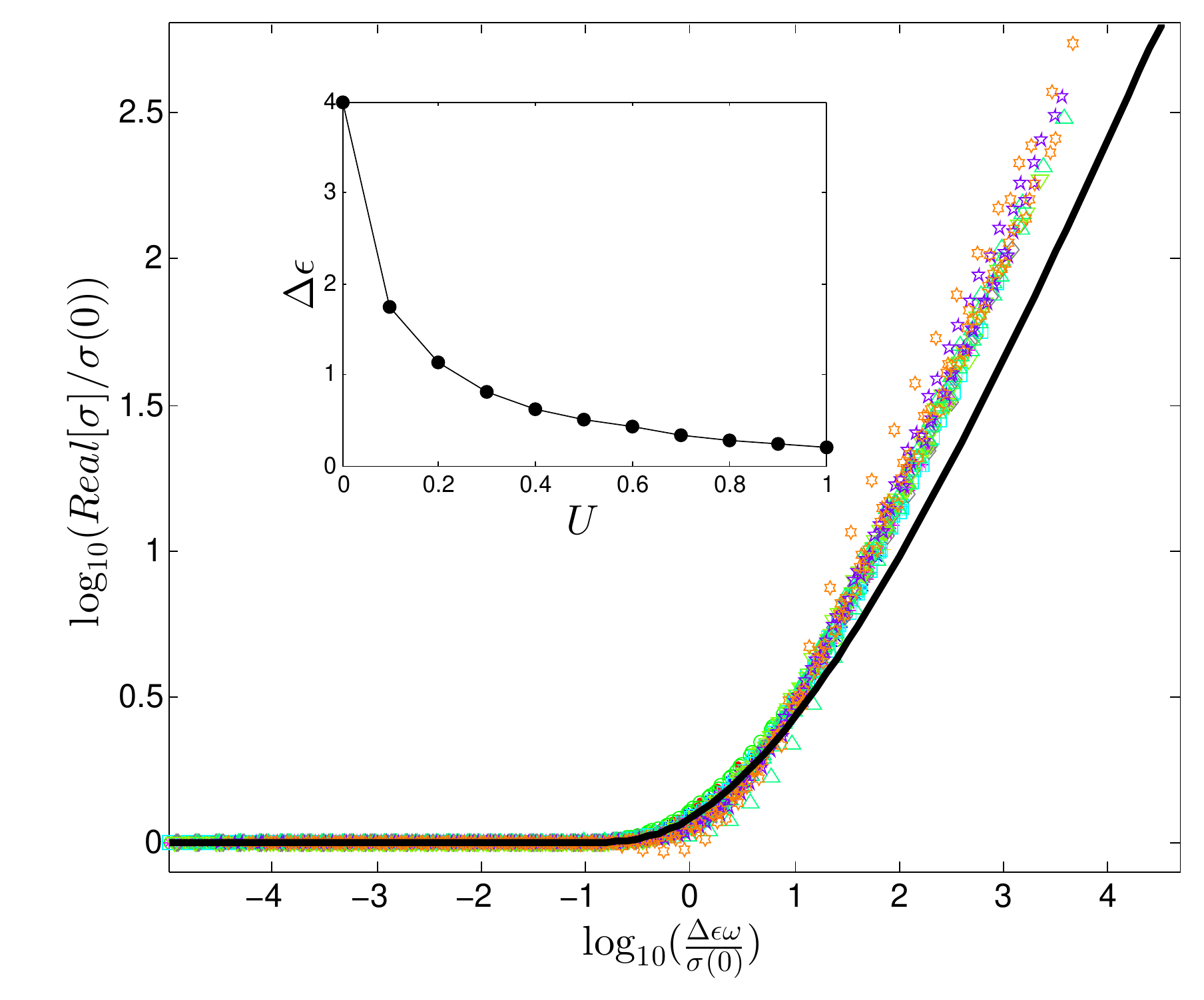}

\caption{(a) The raw numerical data for the real part of the conductance as a function of frequency, obtained by averaging 100 runs for $N=1000$, for a 3d system. Each color and symbol corresponds to a different interaction strength, characterized by the dimensionless parameter $U \equiv \frac{e^2/{r_{nn}}}{W}$, the ratio of the typical nearest-neighbor interaction strength to the width of the quenched disorder. The temperature is $T/W=0.05$. Similar plots are obtained for different values of temperature. (b) The data for the different temperatures and interactions is taken below the saturation frequency, and is scaled using the DC conductance and dielectric constant $\Delta \epsilon$. Data collapse of the numerical data is obtained for all 66 curves (11 values of interaction strengths, and 6 different temperatures for each). For each interaction strength, the temperatures was varied in the range $T/W=0.05$ to $T/W=0.1$. The black solid curve shows the result of the effective medium approximation \cite{review_universal_ac}, which captures the behavior qualitatively.}
\label{res}
\end{figure}

\begin{figure}[h]
\includegraphics[width=0.7\linewidth]{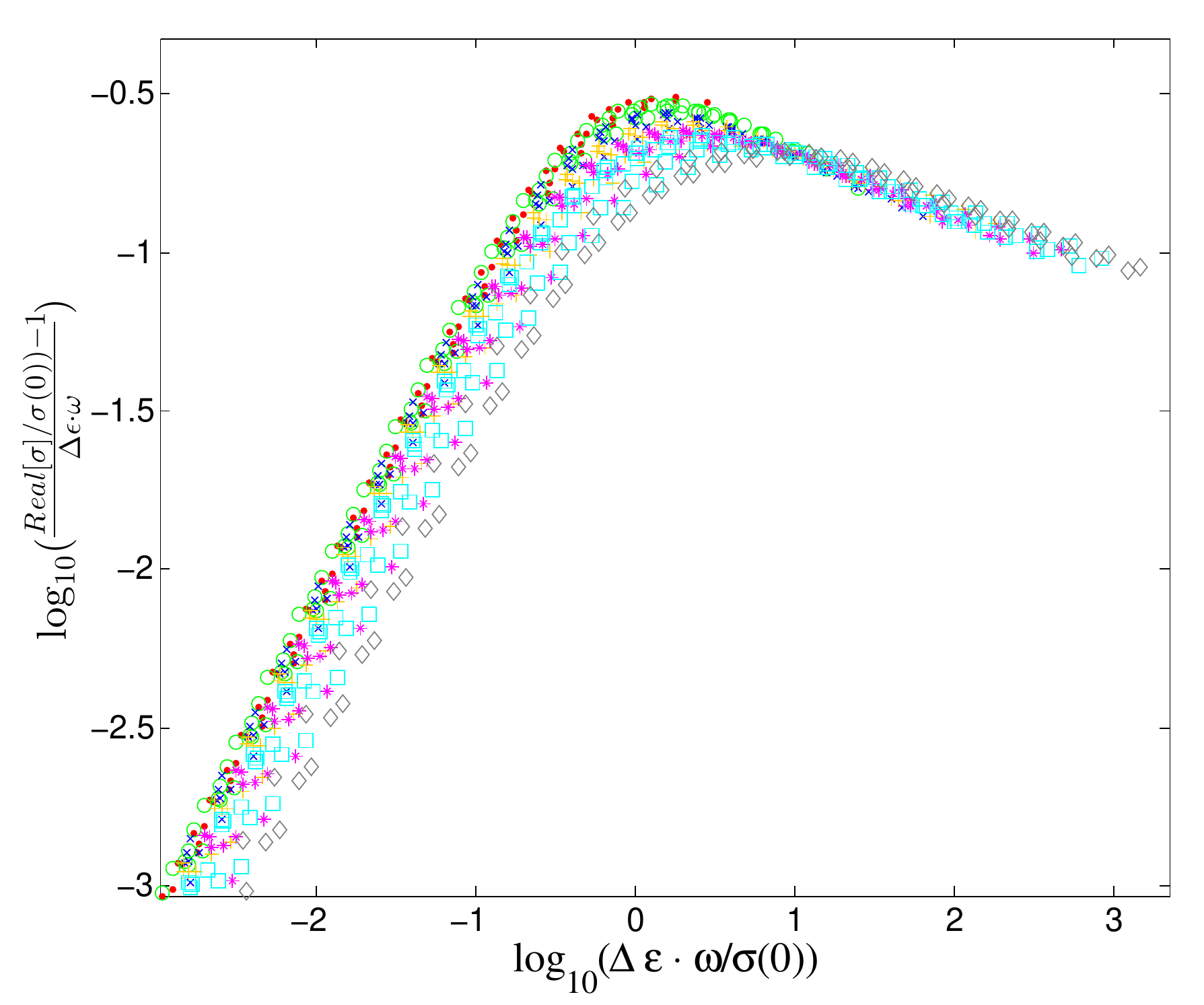}
\caption{The `loss peak' (the maximum in the dissipative component of the dielectric response when plotted as a function of frequency) is plotted when varying temperature and interactions strength, using the scaling of Fig. \ref{res}. The parameters used and color code is as in Fig. \ref{res}.}
\label{loss}
\end{figure}

Fig. \ref{res} shows the real part of the conductance, $\sigma'$, for a 3d system, averaged over 100 realizations for $N=1000$, for different values of temperature and interaction strength. At low-frequencies $\sigma'$ is approximately constant, and the DC value $\sigma(\omega=0)$ corresponds to variable-range-hopping, as expected \cite{amir_vrh}. Upon rescaling both frequency and conductance axis according to the DC conductance (known as Taylor-Isard scaling \cite{taylor1, isard2, review_universal_ac}) the curves do not data collapse.  However, when the frequency axis is further rescaled by the DC dielectric constant $\Delta \epsilon \equiv \rm{lim}_{\omega \rightarrow 0} \frac{\rm{imag}[\sigma(\omega)]}{\omega}$, all plots corresponding to different temperatures and interaction strengths approximately collapse to a \emph{single} curve:
\be \sigma (\omega) = \sigma(0) f[\omega \Delta \epsilon /\sigma(0)], \label{scaling}\ee as shown in Fig. \ref{res}b.
Fig. \ref{loss} shows the scaled loss peak for this scenario, whose maximum occurs roughly at the crossover frequency from the DC to AC regimes.

This form of scaling was first introduced by Sidebottom \cite{sidebottom}, where he also argued for it in terms of the ``Maxwell time" describing current discharge through a medium. Indeed, using the ``RC time" intuition it is plausible that both the resistance matrix (which determines the DC conductance) and the capacitance matrix (which determines the dielectric constant) will affect the relevant timescales. In an elegant argument, it was shown that the rescaling of the frequency axis by $\Delta \epsilon$ is in fact a necessary consequence of the possibility to scale the response at different temperatures \cite{dyre1, review_universal_ac}. The fact that the data scales tells us that the crossover frequency $\omega_c$ from the DC to AC conductance regimes is given by $\omega_c \sim \sigma(0)/\Delta \epsilon$.

It is remarkable that the whole frequency-dependence of the conductance depends only on two parameters associated with the low-frequency limit, namely, the DC conductivity and dielectric constant. The Coulomb interactions are known to lead to the formation of a soft gap in the single particle density-of-states (DOS), the Coulomb gap \cite{efros}. This modifies the temperature behavior of the DC conductance, suppressing it at sufficiently low temperatures \cite{amir_vrh}. The inset of Fig. \ref{res}b shows that the dielectric constant is also significantly suppressed as a result of the interactions. This is plausible since the Coulomb gap will make it harder to find sites with energies close to the Fermi energy, and hence it will be harder to create a polarization in the system.

\emph{Discussion.} The physical basis for the behavior of $\sigma(\omega)$ has been discussed before in the context of effective medium approximations, the pair approximation  and percolation theory \cite{review_universal_ac}. The essential physics underlying the frequency dependent conductance can be illustrated by considering a toy-model one dimensional RC network, consisting of a series of resistors with broadly distributed resistances, each having a capacitor in parallel to it \cite{dyre2}. It can be shown that this simple model obeys the universal scaling, with the value of the capacitor determining $\Delta \epsilon$. Within our approach, the interactions will significantly affect the statistics of the distributions of both resistors and capacitors distribution. The formation of the Coulomb gap due to the interactions broadens the distribution of resistors and capacitances. Nevertheless, by rescaling the frequency and conductance axis by $\sigma(0)$, the DC conduction, and additionally rescaling the frequency axis alone by $\Delta \epsilon$, we can account for the complete frequency response at all frequencies using two quantities, $\Delta \epsilon$ and $\sigma(0)$. The model which we study here provides a tractable case where the effects of temperature and interactions on the frequency dependent response can be understood. The model also manifests a loss-peak, which using the universal form of scaling becomes temperature independent. The lack of temperature dependence of the rescaled plots agrees with experimental results, yet cannot be captured within effective medium approximations and other theoretical approaches \cite{long,isard}. Our model provides a realistic description that agrees with the experimental results, yet is amenable to analytic treatment, with testable predictions such as the dependence of the dielectric constant on interaction strength, and the universal scaling of the conduction. We believe that it would be rewarding to make experimental tests of these results in electron glasses, and see how screening the Coulomb interactions by putting a metal plate close to the sample affects the frequency dependent response.

\emph{Acknowledgments}  This research was supported by the Harvard Society of Fellows and the Milton Fund. The author acknowledges useful discussions with S. Gopalakrishnan, B. I. Halperin, Y. Meroz, Y. Oreg, M. Pollak, B.~I. Shklovskii and B. Skinner.

%

\newpage

\section {Universal frequency-dependent conduction of electron glasses - Supplementary Information}

\subsection {Extracting the conductivity from the steady-state solution}

From Eq. (10) of the main text we can readily find all currents in the system, as well as the currents to the leads. The total conductance $\sigma$ will comprise of two contributions: one from the charge tunneling from the leads into the sample, and another mechanism, effective only at finite frequencies, due to the time dependence of the polarization, which produces a change in the surface charge on the lead and thus a current. Hence:

\be I_{tot} = I_{lead} + \frac{1}{L} \frac{dP}{dt}. \ee

To evaluate the second term, it is useful to use $P= \sum_j \delta n_j x_j$, where $x_j$ is the component of the site position in the direction of the applied field, and $\delta n$ is the (oscillating) change in the occupation number resulting from the sinusoidal voltage. Thus:

\be I_{pol} =\frac{1}{L} \frac{dP}{dt} = \frac{1}{L} \sum_j \frac{\delta n_j}{dt} x_j =  \frac{1}{L} \sum_{j \neq k} I_{kj} x_j. \ee

\subsection {Frequency-dependent conduction of localized circuit elements}

In contrast to the DC conduction which can be described by the flow of current on a percolation cluster \cite{ambegaokar}, at finite frequency even localized circuit components which are disconnected from the rest of it may contribute to the AC current, as shown in Fig. 1b of the main text. Consider a pair of close in space and resonant in energy sites; this pair will significantly contribute to the AC current, since they have a low value of resistance between them, and a high value of capacitance (since their energies are close to the Fermi energy). The impedance associated with this sub-circuit can readily be found by summing the impedance of the capacitors with that of the resistor, leading to: $Z(\omega) = R+ \frac{2}{i \omega C}$. Hence:

\be \sigma_{pair}(\omega) =\frac{R\omega^2 C^2}{R^2\omega^2 C^2 +4}.\ee

Thus, this pair will begin to contribute to the AC conduction at a frequency $\omega_c \sim 1/RC$, and at high frequencies compared with this frequency its contribution will saturate at a value of $\sigma_{max}=1/R$.
%
%

We can account for the frequency dependence at high frequencies analytically, by summing the contributions of many such pairs. For simplicity let us consider $C$ to be constant, and take $R \sim e^{r/\xi}$, where $r$ is the distance between close pairs (i.e., we are ignoring the energy dependence of the resistors, and focusing on the exponential dependence on the tunneling distance). Note that the distribution of distance between close pairs, $P(r)$, has only a polynomial dependence on $r$. Therefore:
\be \sigma(\omega) \approx \int \frac{R\omega^2 C^2 P(R)}{R^2\omega^2 C^2 +4}  dR \approx \int_{r<\xi \log(\omega C)} e^{r/\xi}P(r)dt \sim \omega C . \label{stat}\ee
This is the essence of the result of Ref. [\onlinecite{austin}] which also accounts for the logarithmic corrections to this linear-in-frequency dependence. Note that for a finite system size, we will have a finite number of pairs contributing to Eq. (\ref{stat}), which will lead to a saturation of the frequency dependent conduction; This can be seen in Fig. 2a of the main text.

\end{document}